\RequirePackage{fix-cm}
\documentclass[smallextended]{svjour3}
\smartqed

\usepackage{amsgen}
\usepackage{amssymb}
\usepackage{setstack}
\usepackage{bbold}
\usepackage{mathrsfs}
\usepackage{graphicx}
\usepackage{epsfig}
\usepackage{epstopdf}
\usepackage{cite}

\def\makeheadbox{\relax}
\def\be{\begin{equation}}
\def\ee{\end{equation}}
\def\bea{\begin{eqnarray}}
\def\eea{\end{eqnarray}}

\begin{document}

\title{Cosmological Solutions with Charged Black Holes}

\author{Rashida~Bibi, Timothy~Clifton and Jessie~Durk}

\institute{Rashida~Bibi$^{1,2,3}$, Timothy~Clifton$^1$ and Jessie~Durk$^1$\\ \\
              1. School of Physics and Astronomy, Queen Mary University of London, UK. \\
              2. School of Natural Sciences, National University of Sciences and Technology, Pakistan.\\
              3. Department of Mathematics and Statistics, International Islamic University, Pakistan.\\ \\
              \email{rashida.jahangir@iiu.edu.pk, t.clifton@qmul.ac.uk, j.durk@qmul.ac.uk}
}

\date{}

\maketitle

\begin{abstract}
We consider the problem of constructing cosmological solutions of the Einstein-Maxwell equations that contain multiple charged black holes. By considering the field equations as a set of constraint and evolution equations, we construct exact initial data for $N$ charged black holes on a hypersphere. This corresponds to the maximum of expansion of a cosmological solution, and provides sufficient information for a unique evolution. We then consider the specific example of a universe that contains eight charged black holes, and show that the existence of non-zero electric charge reduces the scale of the cosmological region of the space. These solutions generalize the Majumdar-Papapetrou solutions away from the extremal limit of charged black holes, and provide what we believe to be some of the first relativistic calculations of the effects of electric charge on cosmological backreaction.
\end{abstract}

\newpage

\section{Introduction}

Inhomogeneous cosmological models are an important tool for investigating the foundational assumptions that go into the standard model of cosmology. One such assumption is that all space-times that are {\it statistically} homogeneous and isotropic on large scales should evolve in precisely the same way as the {\it perfectly} homogeneous and isotropic Friedmann-Lema\^{i}tre-Robertson-Walker (FLRW) solutions of Einstein's equations. However, it has never been proven that this assumption should be true under general circumstances, and there are in fact good reasons to question its veracity. This is because averaging and evolution operations do not commute in Einstein's theory, which opens the possibility that small-scale inhomogeneities could affect the large-scale dynamics of the Universe \cite{1,1a,1c}. Inhomogeneous cosmologies that do not rely on averaging are ideal for investigating this backreaction question, and in this paper we will extend a class of such models to include massive bodies with electric charge.

The black hole lattice models that we will be considering here have stimulated a good degree of interest over recent years, having been the subject of investigation by a number of different research groups \cite{PF1,PF2,PF3}, \cite{JL1,JL2}, \cite{VS1,VS2,VS3,VS4}, \cite{RW2,RW3,RW4}, \cite{TC1,TC2,TC3,TC4,TC5,TC6,TC7}, \cite{EB1,EB2,EB3,EB4,EB5}, \cite{CMY0,CMY1,CMY2}. Many of these works build on the pioneering studies performed in refs. \cite{2} and \cite{3}, which provided the Misner initial data for $N$ black holes in asymptotically flat spaces. Some of these approaches have relied on approximation schemes \cite{PF1,PF2,PF3,JL1,JL2,VS1,VS2,VS3,VS4,RW2,RW3,RW4}, while others use the methods of numerical relativity \cite{EB1,EB2,EB3,EB4,EB5,CMY0,CMY1,CMY2}. Most of them have considered vacuum solutions of Einstein's equations, but some have involved generalisations that include a non-zero cosmological constant \cite{VS3,RW3,TC6,CMY2} and the inclusion of matter modelled as radiation and a scalar field \cite{VS3,TC7}. In this paper we focus on including electric charge, so that Reissner-Nordstr{\"o}m-like black holes \cite{stahl}, or other types of charged astrophysical bodies \cite{bibi}, can be included in cosmology. As far as we are aware, this is the first treatment of cosmological backreaction in the presence of non-zero electric charges.

There are several motivations for including electric charge in these models. The first is that it offers a generalisation of existing models, analogous to the generalisation from Schwarzschild to Reissner-Nordstr{\"o}m in the study of static black holes. This should be expected to give new mathematical and physical insights into the nature of these inhomogeneous cosmologies. Moreover, in an FLRW universe the homogeneity of space forbids regions of charged spacetime by definition. There is also relatively little known about the cosmological consequences of electric charge in inhomogeneous models. It therefore remains a somewhat open question as to what the effects of allowing electric charge in cosmology should be in general. The lattice models studied in refs. \cite{PF1,PF2,PF3,JL1,JL2,VS1,VS2,VS3,VS4,RW2,RW3,RW4,TC1,TC2,TC3,TC4,TC5,TC6,TC7,EB1,EB2,EB3,EB4,EB5,CMY0,CMY1,CMY2} allow us the opportunity to study this problem in a well-defined and precise way. Moreover, this work offers the opportunity to generalise the well-known Majumdar-Papapetrou and Kastor-Traschen solutions for multiple extremal black holes with $\vert q \vert =m$ \cite{9,10,KT} to the case where the black holes have $\vert q \vert \neq m$. Finally, the inclusion of electric charge offers a way to break the discrete rotational symmetries that might otherwise exist within these lattice cosmologies, as we find that the black holes must necessarily contain non-identical charges.

This rest of this paper is organised as follows: in Section 2 we review the Einstein-Maxwell constraint equations on a time-symmetric hypersurface. Then in Section 3 we derive initial data for a cosmological model that contains electrically charged black holes. The example of a universe that contains eight black holes is considered in Section 4, which we then compare with a Friedmann cosmology before calculating the apparent horizons of the black holes. Finally we conclude in Section 5. We find that as the magnitude of the electric charge is increased, the discrepancy between the uncharged lattice models and the corresponding FLRW solutions reduces, and that the horizon size of each black hole decreases. Throughout this paper we adopt geometrised units where $8 \pi G = c= 1$.

\section{Geometrostatics with an electric field}
\label{sec:constraint}

The Einstein-Maxwell equations that govern electromagnetic fields in the presence of gravitation are given by
\be
R^\mu_{\phantom{\mu} \nu}-\frac{1}{2} \delta^\mu_{\phantom{\mu}\nu} R = \frac{1}{2} \left( F_{\nu\alpha}F^{\alpha\mu}-\frac{1}{4}\delta^\mu_{\phantom{\mu}\nu} F_{\alpha\beta}F^{\alpha\beta} \right) \equiv T^\mu_{\phantom{\mu} \nu} \, , \label{16}
\ee
where $R_{\mu \nu}$ is the Ricci tensor, and $R$ is the Ricci scalar. The Faraday tensor is denoted by $F_{\mu \nu}$, and obeys the following differential relations:
\be
F^{\mu\nu}_{\phantom{\mu \nu};\nu} = 0 \qquad {\rm and} \qquad  F_{[\mu\nu ; \alpha]}=0 \, ,\label{17}
\ee
where the indices $\mu, \nu, ...$ run from $0$ to $3$. It is these equations that we wish to solve, to obtain the geometry of a universe that contains $N$ electrically-charged black holes.

To find the relevant initial data for this system, we start by choosing a space-like hypersurface with metric $\gamma_{ij}$ and normal $u^{\mu}$, where  $i, j,...$ run from $1$ to $3$. The four constraints from equation (\ref{16}) can then be found to be
\begin{eqnarray}
R^{(3)}+K^2-K_{ij}K^{ij}&=& 2 \rho \, \label{energy} \\[5pt]
\large( K^j_{\phantom{j} i}-\delta^j_{\phantom{j} i} K \large)_{: j}&=& q_i  \label{momentum} \, ,
\end{eqnarray}
where $R^{(3)}$ is the Ricci curvature scalar of the intial 3-dimensional space, $K_{ij}$ is the extrinsic curvature of this 3-space in the 4-dimensional space-time, and where
$$
\rho \equiv T_{\mu \nu} u^{\mu} u^{\nu} \qquad {\rm and} \qquad  q_i \equiv - \delta^j_{\phantom{j} i} u^{\mu} T_{\mu j}
$$
are the energy and momentum densities, respectively. All indices in these expressions should be understood to be raised and lowered with the metric of the 3-space $\gamma_{ij}$, and the covariant derivative symbolized with a colon is with respect to $\gamma_{ij}$.

If our initial hypersurface and matter fields are time-symmetric then we immediately have $K_{ij}=0$, while $F_{i \mu} = - E_i u_{\mu}$ and $F_{ij}=0$. This means that $\rho= E_i E^i$ and $q_i=0$, so that the energy constraint equation (\ref{energy}) becomes
\be
R^{(3)} = 2 E_i E^i \label{init1}
\ee
and the momentum constraint equation (\ref{momentum}) is identically satisfied. Simultaneously, the differential relations from equation (\ref{17}) imply
\be
E^i_{\phantom{i} : i} =0  \label{init2} \, .
\ee
If equations (\ref{init1}) and (\ref{init2}) are both satisfied, then we have a solution to the constraint equations. This provides the initial data for a unique evolution under the six remaining Einstein-Maxwell equations in (\ref{16}) and the three evolution equations in (\ref{17}) \cite{4b}.

Let us now make the following ans{\"a}tze: firstly, let us write the metric as
\begin{equation}
ds^2= \chi^2 \psi^2 h_{ij} dx^i dx^j \, , \label{nbhs}
\end{equation}
where $\psi$ and $\chi$ are functions of all spatial coordinates, and where $h_{ij}$ is the metric of a conformal 3-dimensional space of curvature $\mathcal{R}= \,\,\, $constant. Secondly, let us choose, as in \cite{2},
\be
E_i = [ \ln \left(\chi / \psi \right) ]_{, i} \, .
\ee
The time-symmetric constraint equations (\ref{init1}) and (\ref{init2}) then become
\be
\label{helm}
\bar{\nabla}^2 \chi = \frac{\mathcal{R}}{8} \chi \qquad {\rm and} \qquad \bar{\nabla}^2 \psi = \frac{\mathcal{R}}{8} \psi \, ,
\ee
where $\bar{\nabla}^2$ indicates a covariant Laplacian with respect to the metric $h_{ij}$. These are two copies of Helmholtz equation, for the two functions $\chi$ and $\psi$. The solutions to this equation are well known, and it can be seen that $E_i \neq 0$ as long as $\chi$ and $\psi$ are not directly proportional. The reader may note that for periodic lattices of black holes the equations in (\ref{helm}) can have positive energy solutions only if $\mathcal{R}$ is positive (see \cite{EB1} for proof).

\section{A universe full of charged black holes}
\label{sec:many}

To find cosmologically interesting configurations of black holes, it is convenient to choose $h_{ij}$ to be the metric of a 3-sphere:
\be
d\bar{s}^2 = h_{ij} dx^i dx^j = d r^2 + \sin^2 r \left( d\theta^2 + \sin^2 \theta \, d\phi^2\right) \, , \label{3sphere}
\ee
so that $\mathcal{R}=6$. By noting that equations (\ref{helm}) are both linear, and by recognising that $f\propto 1/\sin(r/2)$ is a solution, it can be seen that we can write solutions to these equations as follows:
\begin{eqnarray}
\chi&=&\sum_{i=1}^{N}\frac{c_i}{2\sin({r_i}/{2})}, \label{q11}\\
\psi&=&\sum_{i=1}^{N}\frac{d_i}{2\sin({r_i}/{2})}, \label{q12}
\end{eqnarray}
where $i$ runs from $1$ to $N$, and where $c_i$ and $d_i$ are two sets of $N$ constants (yet to be determined). These solutions contain $N$ poles on the conformal 3-sphere, located at arbitrarily chosen locations, and the symbol $r_i$ is intended to denote the value of the $r$ coordinate after rotating the sphere so that the $i$th pole is located at $r_i=0$. Each of these poles will correspond to an electrically-charged mass point, at the centre of a black hole in a cosmological model.

\subsection{A slice through Reissner-Nordstr{\"o}m}

In principle, equations (\ref{q11}) and (\ref{q12}) represent exact initial data for $N$ electrically charged black holes in a hyperspherical cosmology. In practise, we must relate the constants $c_i$ and $d_i$ to the charge and mass of each of the black holes, in order for the geometry to be fully specified. To make this connection, it is useful to recall that the Reissner-Nordstr{\"o}m solution with mass $m$ and charge $q$ can be written as
\be
ds^2=-\left( 1-\frac{2m}{\sigma}+\frac{q^2}{\sigma^2} \right) dt^2+\frac{d\sigma^2}{\left( 1-\frac{2m}{\sigma}+\frac{q^2}{\sigma^2} \right)} +\sigma^2 ( d\theta^2+\sin^2\theta d\phi^2 ). \label{m1}
\ee
By making the transformation
\begin{equation}
\sigma=\rho \left( 1+\frac{m}{2 \rho} \right)^2- \frac{q^2}{4 \rho} \, ,
\end{equation}
we can therefore write a time-symmetric slice of the metric from equation (\ref{m1}) as
 \begin{equation}
ds^2= \left( 1+\frac{m-q}{2\rho} \right)^2 \left( 1+\frac{m+q}{2\rho} \right)^2 \left( d\rho^2+\rho^2d\theta^2+\rho^2\sin^2\theta d\phi^2 \right) \, .
\end{equation}
A further transformation, to cast this metric in the form of equations (\ref{nbhs}), (\ref{q11}) and (\ref{q12}), is given by $\rho=k\tan({r}/{2})$, where $k$ is a constant. This transforms the conformal 3-space from a plane to a sphere, and results in
\begin{eqnarray}
\label{keq}
ds^2=\frac{k^2}{4} \left( \frac{1}{\cos({r}/{2})}+\frac{m-q}{2k \sin({r}/{2})} \right)^2 \left( \frac{1}{\cos({r}/{2})}+\frac{m+q}{2k\sin({r}/{2})} \right)^2d\bar{s}^2 \, , \label{m110}
\end{eqnarray}
where $d\bar{s}^2$ is the line-element on a 3-sphere, as given in equation (\ref{3sphere}). Finally, if we choose $2 k = \sqrt{m-q}\sqrt{m+q}$, then we get
\begin{equation}
ds^2= \left(\frac{\sqrt{m+q}}{2\sin({r}/{2})}+ \frac{\sqrt{m-q}}{2\cos({r}/{2})} \right)^2 \left(\frac{\sqrt{m-q}}{2\sin({r}/{2})}+\frac{\sqrt{m+q}}{2\cos({r}/{2})} \right)^2 d\bar{s}^2   \, .\label{m11}
\end{equation}
This result is of exactly the form that we would have obtained from considering equations (\ref{nbhs}), (\ref{q11}) and (\ref{q12}) with one mass positioned at $r=0$ and a second mass positioned at $r=\pi$. If we further choose $c_1=d_2=\sqrt{m+q}$ and $c_2=d_1=\sqrt{m-q}$ then the two geometries are formally identical. This means that the time-symmetric initial data for the Reissner-Nordstr{\"o}m solution can be considered as a special case of our more general intial data, with one mass of charge $+q$ at $r=0$ and a second mass of charge $-q$ at $r=\pi$.

\subsection{$N$ arbitrarily positioned, charged black holes}

Let us now consider the general case, where $N$ masses are positioned at arbitrarily chosen locations. In the vicinity of a mass point, and by using equation (\ref{keq}), we can relate $c_i$ and $d_i$ to the mass $m_i$ of each of the black holes in any arbitrary distribution of points. The first step in this is rotating coordinates so that the $i$th mass appears at $r=0$. In the limit when $r \rightarrow 0$ we can then expand the terms in equation (\ref{nbhs}) to find
\begin{equation}
ds^2=\left(\frac{c_i}{r}+\sum_{j\neq i}\frac{c_i}{2\sin (r_{ij}/2)} \right)^2 \left(\frac{d_i}{r}+\sum_{j\neq i}\frac{d_i}{2\sin( r_{ij}/2)} \right)^2 d\bar{s}^2 \, , \label{d1}
\end{equation}
where $r_{ij}$ is the coordinate distance between the positions of the $i$th and $j$th masses. Comparing this equation with the expanded version of equation (\ref{m110}), again around $r=0$, allows us to read off the following from the coefficients of the leading-order terms:
\be
k = \frac{1}{2 c_i d_i} \left( c_i \sum_{j\neq i}\frac{d_j}{2 \sin(r_{ij}/2)} +d_i  \sum_{j\neq i}\frac{c_j}{2 \sin(r_{ij}/2)} \right) - \frac{q_i^2}{2 c_i d_i}
\ee
and
\begin{equation}
m_i=c_i\sum_{j\neq i}\frac{d_j}{2\sin(r_{ij}/2)}+d_i\sum_{i\neq j}\frac{c_j}{2\sin(r_{ij}/2)} \, .\label{mi}
\end{equation}
The former of these equations simply relates the coordinate systems of the Reissner-Nordstr{\"o}m solution and our multi-black hole solution in the vicinity of one of our mass points. The latter, however, allows the mass of each of our black holes to be identified. That is, if a multi-black hole solution is specified (with a full set of $c_i$s and $d_i$s), then the above analysis shows that the geometry of space in the vicinity of any one of the black holes will be similar to a Reissner-Nordstr{\"o}m black hole with mass $m$, where $m$ is given by equation (\ref{mi}).

It now remains to identify the electric charge on each of our black holes. A general definition of electric charge within a region $\Omega$ can be given by \cite{2}
\begin{equation}
q_i \equiv \frac{1}{4\pi}\int_{\partial \Omega}{E_in^idS} \, ,\label{q1}
\end{equation}
where $E_i$ is the electric field and $n^i$ is the unit inward-pointing normal. For our black holes, it is convenient to take the boundary $\partial \Omega$ to correspond to asymptotic infinity, on the far side of the Einstein-Rosen bridge. This gives
\begin{equation}
n^i = \left( \frac{-1}{\psi\chi},0,0 \right) \, . \label{q2}
\end{equation}
and $dS= \psi^2 \chi^2 r^2 \sin\theta d\theta d\phi$, where in both of these expressions $\chi$ and $\psi$ should be taken to be evaluated in the limit $r \rightarrow 0$. Evaluating equation (\ref{q1}) in this limit, and using equations (\ref{q11}), (\ref{q12}) and (\ref{q2}), gives
\begin{equation}
q_i=c_i\sum_{j\neq i}\frac{d_j}{2\sin(r_{ij}/2)}-d_i\sum_{j\neq i}\frac{c_j}{2\sin(r_{ij}/2)} \, . \label{qii}
\end{equation}
This equation gives the charge on the mass located at $r_i=0$, and can equally be used to evaluate the charge on every other black hole in our solution. It is interesting to note that equation (\ref{qii}) immediately gives $\sum_i q_i=0$, meaning that the total charge in the universe must be zero, independently of how the black holes are distributed and their masses. This makes sense physically, as lines of flux can only end on masses in a closed cosmology.

Adding and subtracting the equations (\ref{mi}) and (\ref{qii}) gives a somewhat simpler pair of equations:
\begin{eqnarray}
m_i+q_i&=&c_i\sum_{j\neq i}\frac{d_j}{\sin(r_{ij}/2)} \label{mi1} \, ,\\[5pt]
m_i-q_i&=&d_i\sum_{j\neq i}\frac{c_j}{\sin(r_{ij}/2)} \label{mi2} \, .
\end{eqnarray}
While relatively simple to write down in this form, this is still a system of non-linear equations, which makes it difficult to solve for the $\{c_i \}$ and $\{ d_i \}$ directly, after the desired $\{m_i \}$ and $\{ q_i \}$ have been specified.

\section{A periodic universe with eight charged black holes}
\label{sec:eight}

Probably the simplest solution to equations (\ref{mi1}) and (\ref{mi2}) is the case where all but one of the black holes is extremal, with $m_i=q_i$. If this is the case, and the one exceptional black hole is labelled $i=1$, then vanishing total charge means that we must have $q_1 = -\sum_{\i \neq 1} q_i = -\sum_{\i \neq 1} m_i$. The only solutions to equations (\ref{mi1}) and (\ref{mi2}) then have $m_1=-q_1$, and we have the initial data for a Majumdar-Papapetrou \cite{9,10} solution with spatial infinity transformed into the black hole with $i=1$. In the next section we will present an example with non-extremal black holes, in which an exact solution for $\{c_i \}$ and $\{ d_i \}$ can be found once the mass and charge of each of the black holes has been specified.

\subsection{Determining the sets of constants $\{c_i \}$ and $\{ d_i \}$}

To find a regular arrangement of black holes in a closed cosmological model, we can choose to tile the conformal hypersphere from our 3-dimensional initial data with eight cubic lattice cells. This is one of the six possible tilings of a 3-sphere with regular convex polyhedra, and corresponds to the structure one would obtain by placing a hypercube within the hypersphere in a 4-dimensional Euclidean embedding space, and by ensuring that the vertices of the hypercube are all touching the hypersphere simultaneously (i.e. by circumscribing the cube with the sphere). The lines that connect the points where these two structures touch then form the edges of eight equal-sized cubes, which can be used as the primitive lattice cells of our tiling. If we place one mass point at the centre of each of these cells, then we have a completely regular lattice in which each mass is exactly equidistant from each of its nearest neighbours.

One of the nice features of the regular eight-mass universe just described is that each of the black holes will be antipodal to another black hole, just as in the time-symmetric slice through the Reissner-Nordstr{\"o}m solution discussed in the previous section. This can be verified by looking at the final column of Table \ref{eight}, which gives the coordinates of the location of each mass in both a set of Cartesian coordinates in the 4-dimensional Euclidean embedding space, as well as in a set of hyperspherical polar coodinates intrinsic to the 3-space itself. Taking a hint from the existence of the time-symmetric Reissner-Nordstr{\"o}m geometry, we can now choose the $\{c_i\}$ and $\{d_i\}$ of these eight masses such that the $c_i$ from each mass is equal to the $d_i$ of the antipodal mass, i.e. so that
\begin{eqnarray}
&&c_1=c_3=c_5=c_7=d_2=d_4=d_6=d_8=e-f \, ,\label{c1}\\[5pt]
&&c_2=c_4=c_6=c_8=d_1=d_3=d_5=d_7=e+f \, ,\label{c2}
\end{eqnarray}
where $e$ and $f$ are constants (yet to be determined). This choice then very conveniently allows us to determine from equations (\ref{mi1}) and (\ref{mi2}) that
\begin{eqnarray}
m_1=m_2=m_3=m_4=m_5=m_6=m_7=m_8 &=& (1+6 \sqrt{2}) e^2 + f^2 \, ,\label{m1a}\\[5pt]
q_1=-q_2=q_3=-q_4=q_5=-q_6=q_7=-q_8 &=& -2 (1+3 \sqrt{2}) e f \, ,\label{m1b}
\end{eqnarray}
so that every black hole in the universe has an identical mass to all of the others, while having an equal and opposite electric charge to its antipodal partner. This seems to be the simplest way to satisfy the requirement that the total electric charge of all black holes must vanish, whilst maintaining at least some degree of regularity.

\begin{table}
\begin{center}
\begin{tabular}{|c|c|c|}\hline
mass number, $i$ & $(w, x, y, z)$ & ($r$, $\theta$, $\phi$) \\ \hline \hline
$1$ & (1, 0, 0, 0) & (0, $\frac{\pi}{2}$, $\frac{\pi}{2}$)\\ \hline
$2$ & (-1, 0, 0, 0) & ($\pi$, $\frac{\pi}{2}$, $\frac{\pi}{2}$)\\ \hline
$3$ & (0, 1, 0, 0) & ($\frac{\pi}{2}$, 0, $\frac{\pi}{2}$)\\ \hline
$4$ & (0, -1, 0, 0) & ($\frac{\pi}{2}$, $\pi$, $\frac{\pi}{2}$)\\ \hline
$5$ & (0, 0, 1, 0) & ($ \frac{\pi}{2}$, $\frac{\pi}{2}$, 0)\\ \hline
$6$ & (0, 0, -1, 0) & ($\frac{\pi}{2}$, $\frac{\pi}{2}$, $\pi$)\\ \hline
$7$ & (0, 0, 0, 1) & ($\frac{\pi}{2}$, $\frac{\pi}{2}$, $\frac{\pi}{2}$)\\ \hline
$8$ & (0, 0, 0, -1) & ($\frac{\pi}{2}$, $\frac{\pi}{2}$, $\frac{3\pi}{2}$)\\ \hline
\end{tabular}
\caption{The coordinate positions of each of the eight black holes in our regular tiling of the conformal 3-sphere. The coordinates $\{w,x,y,z\}$ are Cartesian coordinates in a 4-dimensional Euclidean embedding space, and the coordinates $\{ r, \theta , \phi \}$ are the hyperspherical coordinates intrinsic to the initial hypersurface, as used in equation (\ref{3sphere}).}
\label{eight}
\end{center}
\end{table}

Finally, it is now straightforward to solve equations (\ref{m1a}) and (\ref{m1b}) for $e$ and $f$ to get
\begin{eqnarray}
e&=&\frac{\sqrt{(19 + 6 \sqrt{2}) m + \sqrt{(433 + 228 \sqrt{2}) m^2 - (91 + 120 \sqrt{2}) q^2}}}{\sqrt{(182 + 240 \sqrt{2})}}
\end{eqnarray}
and $f=-{q}/{2 (1+3 \sqrt{2}) e}$, where $m=m_i$ is the mass of each black hole and $\vert q \vert=\vert q_i \vert$ is the magnitude of the charge on each black hole (with sign chosen so that $q=q_1$). We now have an explicit solution where all black holes have equal mass, and half of them have positive charge while the other half have negative charge.

\subsection{Comparison with Friedmann cosmology}

With a cosmological solution in hand, it is interesting to check whether the large-scale properties of our space bear any resemblance to the predictions of the commonly used homogeneous and isotropic FLRW models. The first thing that one could compare, between these two cases, is the scale of the cosmological region at the maximum of expansion. The initial data constructed above can immediately be seen to be at such a maximum, as it was chosen to be time-symmetric. The scale factor at the corresponding moment of time in a spatially closed and dust dominated FLRW model, which we take to be the most similar to our configuration of black holes, is given by
\begin{equation} \label{aFLRW}
a(t_{\rm max}) = \frac{4 M}{3 \pi} \, ,
\end{equation}
where $M$ is the total mass in dust in the entire space-time. By comparing this number to the scale of the cosmological region in our black hole space-time we can then deduce the effect of condensing the mass into a finite number of points, with discretely distributed electric charges. This will extend previous results concerning cosmologies filled with uncharged black holes, and will allow us to consider the cosmological effects of electrical charge.

The most natural way to determine the scale of the cosmological region of the black hole space-time is to calculate the proper length of one of the edges of one of our lattice cells. If we rotate our solution until a cell edge lies along a curve with $\theta$ = constant and $\phi$ = constant, then the proper length of the edge is given by
\begin{equation}
d=\int_{r_1}^{r_2}\sqrt{g_{rr}} dr =\int_{r_1}^{r_2} \psi\chi dr \, , \label{d1}
\end{equation}
where $r_1$ and $r_2$ are the coordinates of the end-points of the edge being studied (i.e. the location of the vertices of the lattice cell). The proper length of a curve that subtends the same angle in a closed FLRW model is given by
\begin{equation}
d_{\rm FLRW}=\int_{r_1}^{r_2}a(t_{\rm max}) dr =  (r_2-r_1) a(t_{\rm max})  \, , \label{d2}
\end{equation}
where $a(t_{\rm max})$ is the scale factor from equation (\ref{aFLRW}). In general, the value of $d$ will depend on both the mass of the black holes and their charge ($m$ and $q$), while the value of $d_{\rm FLRW}$ will depend only on the total mass in the space $M$ (once the locations of the cell vertices have been chosen). If we take $M=8 m$, so that each cosmology has the same total proper mass, then the relevant comparison of scales in the two cosmologies will be given by
\begin{equation}
\frac{a}{a_{\rm FLRW}} = \frac{d}{d_{\rm FLRW}} \, ,
\end{equation}
where $d$ and $d_{\rm FLRW}$ are given by equations (\ref{d1}) and (\ref{d2}). This ratio is only a function of the charge to mass ratio of the black holes $q/m$, and reduces to the measure of cosmological back-reaction studied previously in ref. \cite{TC1} in the limit $q\rightarrow 0$. The reader may note that this comparison between FLRW and our black hole lattice neglects the contribution of interaction energies between black holes, which may be considerable. A study of the effect of these quantities on the global cosmology will appear elsewhere \cite{new}.

\begin{figure}[t!]
\begin{center}
  \includegraphics[width=11cm]{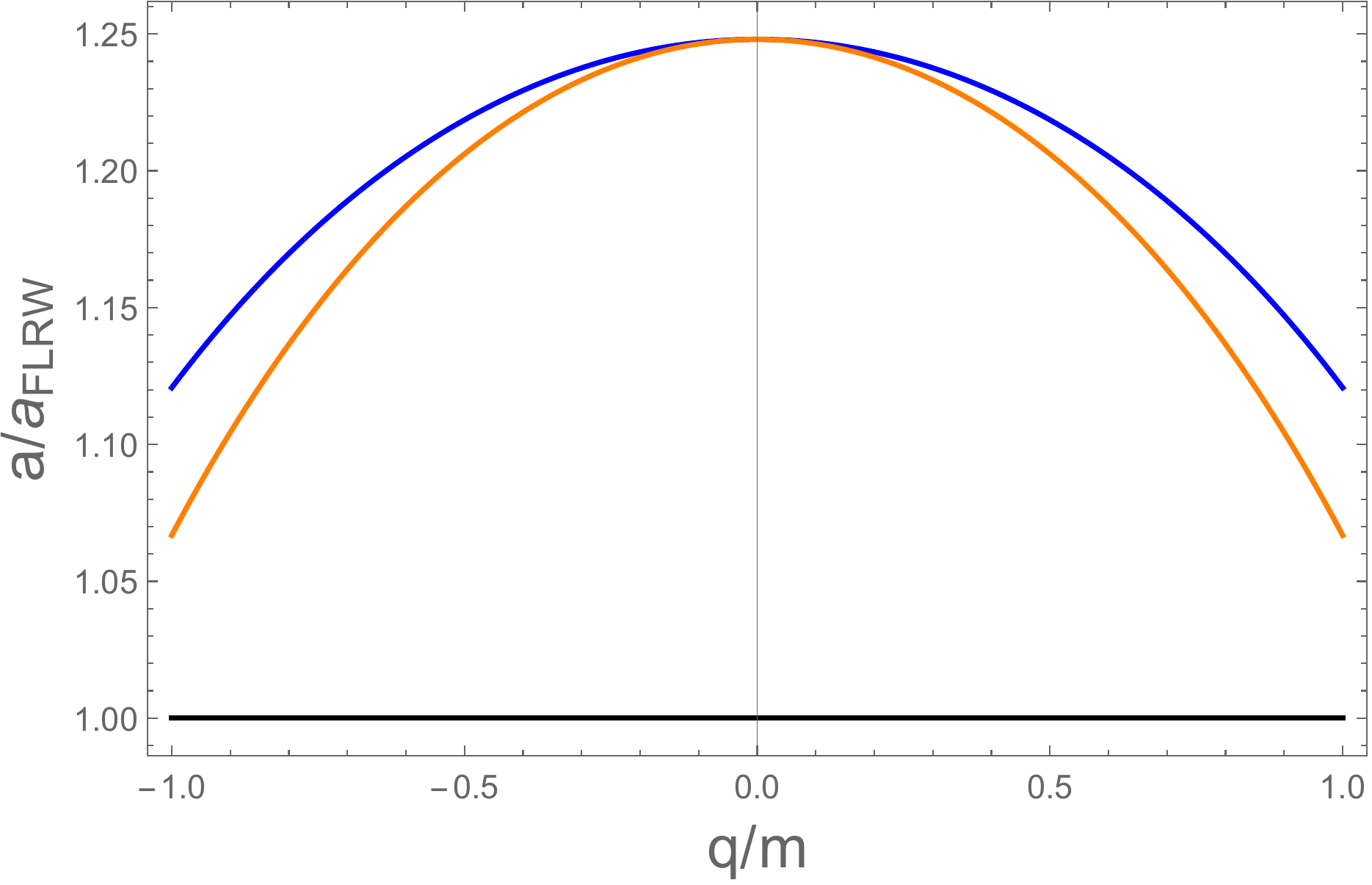}
  \caption{The scale of the cosmological region in our black hole lattice, compared to a spatially closed dust-dominated FLRW model with the same mass at the maximum of its expansion. The two lines correspond to measuring the scale of the black hole universe either along a curve that joins cells that contain three black holes with the same charge (orange), or two black holes with the same charge and a third with the opposite charge (blue).}  \label{1.4200}
  \end{center}
\end{figure}

The value of the ratio of scale factors as a function of $q/m$ is straightforward to calculate, and is displayed graphically in figure \ref{1.4200}. When $q/m \rightarrow 0$ we recover the results of the uncharged case, and as $q/m \rightarrow \pm 1$ the black holes become extremal. It can be seen that the difference from the predicted scale factor from FLRW cosmology is greatest when the black holes are uncharged. Increasing the charges on the black holes decreases the discrepancy with FLRW, but the scale of the cosmological region in our black hole cosmology is always greater than that of the corresponding FLRW model, even when the black holes approach extremality. The reader may note that there are two curves in figure \ref{1.4200}. These correspond to the situations where the three cells that meet along the cell edge under investigation all contain black holes that have the same charge (orange line), or when two of the cells have the same charge and the other has the opposite charge (blue line). The results do not depend on whether the same charges in these two situations are positive or negative; the ratio of scale factors is the same in either case. These two curves therefore describe every cell edge that exists within our eight-black hole model.

As well as the scale of the cosmological region in our black hole universe, it is also of interest to determine the location of the horizons around each of the black holes. This is partly to ensure that there is no overlap in the horizons corresponding to different black holes, so that the solution can reasonably be referred to as a ``cosmological model''. For our purposes, the location of the event horizon is impossible to locate, as we do not have knowledge of the global space-time. The position of the apparent horizon, however, can be determined from knowledge of the initial data alone. We will therefore use this surface to approximate the location of the horizon of each of our black holes.

The apparent horizon is the outermost marginally outer trapped surface, by definition. As our initial data is time-symmetric, this surface must be an extremal surface in this 3-space \cite{11}. We can approximate its location by rotating our 3-sphere until one of the masses appears at $r=0$. The area of any sphere centered on this mass must then be given by
\begin{equation}
A=\int^{2 \pi}_0 \int^{\pi}_0 \psi^2\chi^2\sin^2{r} \sin{\theta} \, d\theta \, d\phi \, ,
\end{equation}
where $r$ = constant. Integrating this quantity numerically, and determining the value of $r$ that minimizes it, then gives us a good approximation to the location of the black hole horizon. This method of assuming a sphere of constant coordinate radius to approximate the apparent horizon necessarily produces a slight overestimate of the horizon area, as the true area of the apparent horizon is that of a minimal sphere in the initial geometry. However, as long as the horizons are well separated, this error is expected to be very small \cite{TC2}.

\begin{figure}[t!]
\begin{center}
\includegraphics[width=11cm]{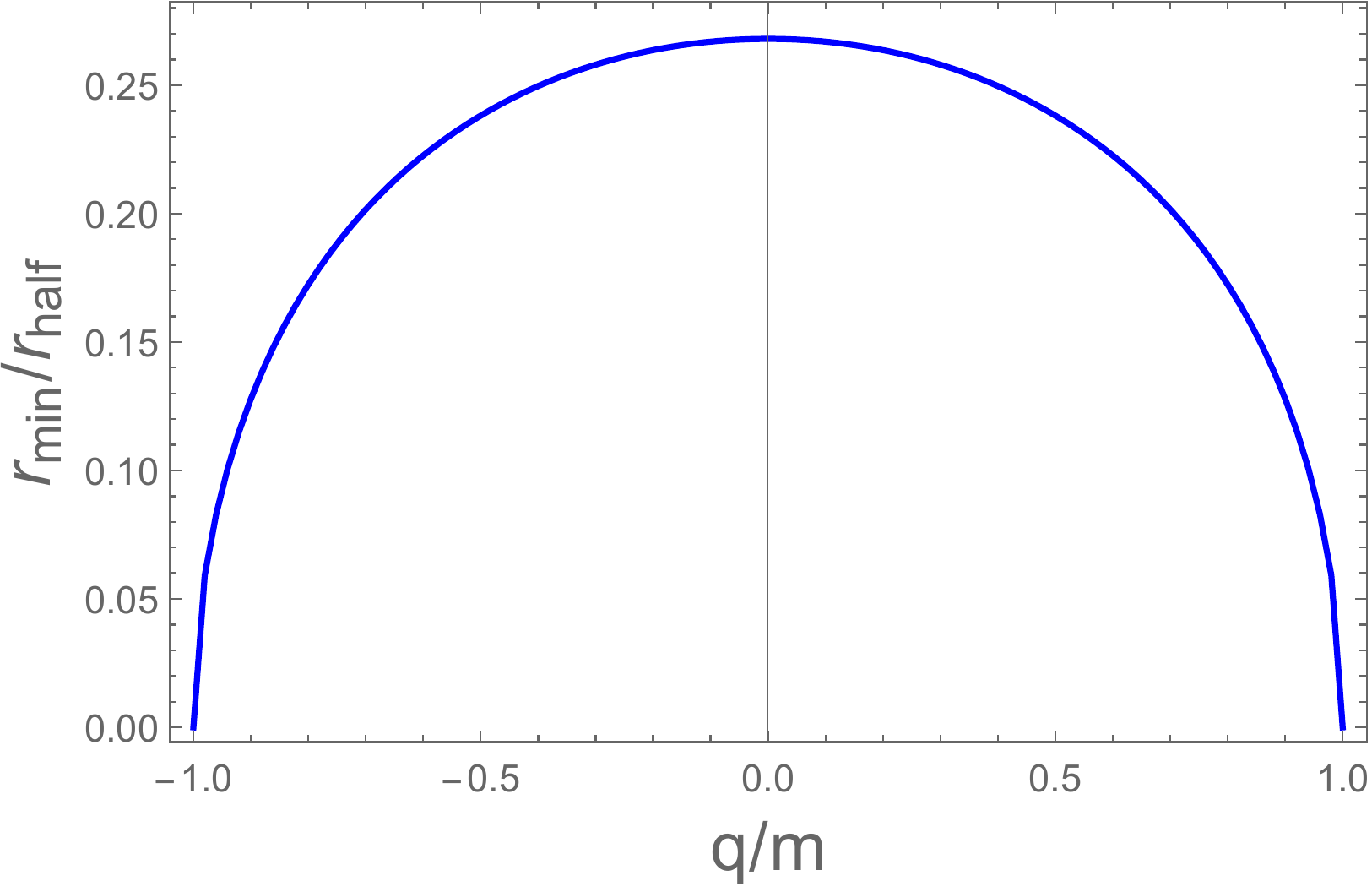}
\caption{The position of the apparent horizon around one of our black holes, $r_{\rm min}$, displayed as a fraction of the coordinate distance to the halfway point between black holes, $r_{\rm half}$. We find $r_{\rm min}\rightarrow 0$ as $q/m \rightarrow \pm 1$.}  \label{horizonplot}
\end{center}
\end{figure}

The results of calculating the position of the apparent horizon in this way are shown in figure \ref{horizonplot}, where the $r$ coordinate of the apparent horizon as a fraction of the coordinate distance between neighbouring black holes is displayed. When $q/m \rightarrow 0$ it can be seen that the horizon extends about $27\%$ of the way to the halfway point between black holes, as expected from previous work. As the charge on the black hole is increased, to either positive or negative values, the horizon then shrinks back towards the centre of the black hole, until it reaches zero in the limit in which $q/m \rightarrow \pm 1$. This is similar behaviour to what should be expected from a maximal slice through a Reissner-Nordstr{\"o}m black hole, and verifies that the black hole horizons in our eight-mass model never touch.

\section{Conclusion}
\label{sec:conc}

We have presented an analysis of an inhomogeneous cosmological model that consists of a lattice of regularly arranged, charged black holes at maximum of expansion. This generalises existing work in the literature and provides insight into the effect that electric charge has on cosmological physics. We have found that the only solutions that exist are those that have zero net charge. We then applied this analysis to a specific model containing eight black holes, and found a relatively simply way of obtaining an uncharged universe overall -- by demanding that black holes exist in antipodal pairs, and that each black hole has equal and opposite charge to its antipodal partner.

We then compared this model to an FLRW model with positive spatial curvature containing the same total mass in dust. We did this by comparing the scale between the two types of universe at maximum of expansion as a function of the charge to mass ratio of the black holes. We found that as the magnitude of the charge on the black holes was increased, the discrepancy between lattice and FLRW was reduced, with the largest discrepancy corresponding to the case where there is no charge. Lastly we investigated the size of the apparent horizon in this initial data, as a function of this charge to mass ratio. We found that as the magnitude of the charge of the black holes is increased the size of the apparent horizon decreases, and that in the limit where the black hole charges become extremal the apparent horizons recede to zero radius.

\vspace{10pt}
\noindent {\bf Acknowledgements:}  RB aknowledges support from the Higher Education Commission (HEC) of Pakistan, which provided funding to visit QMUL while this work was performed. TC and JD are supported by the STFC. We would also like to thank the late Bill Bonnor for suggesting cosmologies with charged black holes as a subject for research.

\end{document}